\documentclass[12pt]{article}
\usepackage{mydef}
\usepackage[show]{ed}
\usepackage{authblk}
\usepackage{url}
\usepackage{algorithm}

\newtheorem{proof}{Proof}

\newtheorem{theorem}{Theorem}[section]
\newtheorem{proposition}[theorem]{Proposition}

\begin{document}

\title{Pricing Illiquid Options with $N+1$ Liquid Proxies Using Mixed Dynamic-Static Hedging
\thanks{Opinions expressed in this paper are those of the authors, and do not necessarily reflect the view of JPMorgan Chase and Numerix}}
\author[1,2]{Igor Halperin}
\author[3,2]{Andrey Itkin}

\affil[1]{{\small MR\&D, JPMorgan Chase, 270 Park Avenue, New York, NY 10172, USA
igor.halperin@jpmorgan.com}}
\affil[2]{\small Polytechnic Institute of New York University, 6 Metro Tech Center, RH 517E, Brooklyn NY 11201, USA}
\affil[3]{\small Numerix LLC, 150 East 42nd Street, 15th Floor, New York, NY 10017, USA
aitkin@numerix.edu}

\date{(submitted to IFTAF)}

\maketitle

\begin{abstract}
We study the problem of optimal pricing and hedging of a European option written on an illiquid asset $Z$ using a set of proxies: a liquid asset $S$, and  $N$ liquid European options $P_i$,
each written on a liquid asset $Y_i, i=1,N$. We assume that the $S$-hedge is dynamic while the multi-name $Y$-hedge is static. Using the indifference pricing approach with an exponential utility,
we derive a HJB equation for the value function, and build an efficient numerical
algorithm. The latter is based on several changes of variables, a splitting scheme, and
a set of Fast Gauss Transforms (FGT), which turns out to be more efficient in terms of
complexity and lower local
space error than a finite-difference method. While in this paper we apply our framework to an incomplete market version of the credit-equity Merton's model, the same approach can be used for other asset classes (equity, commodity, FX, etc.), e.g. for pricing and hedging options with illiquid strikes or illiquid exotic options.
%Calibration of the model is discussed as well as typical market values of the risk aversion parameter $\gamma$.
\end{abstract}

%\keywords{incomplete market, indifference pricing, illiquid option, dynamic/static hedge, HJB equation, splitting}

%==================================================
\section{Introduction}
%==================================================
This work is an extension of our paper \cite{HalperinItkin2012} where the following problem was considered. We support a trader who wants to buy (or sell) a European option $ C_Z $ on asset $Z$ with maturity $ T $ and payoff $ G_Z $. The trader wants to hedge this position, but the underlying asset $Z$ is illiquid. However, some liquid proxies of $Z$ are available in the marketplace. First, there is a financial index (or simply an {\it index})  $S$ (such as e.g. S\&P500 or CDX.NA)\footnote{Here we refer to this instrument as an index, but it could be any "linear" instrument such as stock, forward, etc.}  whose market price is correlated with  $Z$. In addition, there is another correlated asset $Y$ which has a liquidly traded option $C_Y$ with a payoff $G_Y$ similar to that of $C_Z$, and with the same maturity $T$. The market price $p_Y$ of $C_Y$ is also known.

Our trader realizes that hedging $Z$-derivative with the index $S$ alone may not be sufficient for a number of reasons. First, she might be faced with a situation where correlation coefficients $\rho_{yz}, \rho_{sz}$ (which for simplicity are assumed to be constant) are such that $\rho_{yz} > \rho_{sz}$. In this case we would intuitively expect a better hedge produced by using $Y$ or $C_Y$ as the hedging instruments. Second, if we bear in mind a stochastic volatility-type dynamics for $Z$, the stochastic volatility process may be "unspanned", i.e. the volatility risk of the option may not be traded away by hedging in option's underlying\footnote{The notion of unspanned stochastic volatility was introduced in \cite{CDG}. For a discussion of such scenarios for e.g. commodities markets, see \cite{TrolleSchwartz2009}.}. If that is the case, one might want to hedge the unspanned stochastic volatility by trading in a "similar" option on the proxy asset $Y$. So our trader is contemplating a hedging strategy that would use both $ S $ and $ Y $. To capture an "unspanned" stochastic volatility, the trader wants to use a derivative $ C_Y $ written on $ Y $ rather than asset $ Y $ directly.

As transaction costs are usually substantially higher for options than for underlyings, our trader sets up a static hedge in $ C_Y$ and a dynamic hedge in $ S_t $. The static hedging strategy amounts to selling (or buying) $ \alpha $ units of $ C_Y$ options at time $ t = 0 $. An {\it optimal} hedging strategy would be composed of a pair $ (\alpha^*, \pi_s^*) $ where $ \alpha^* $ is the optimal static hedge, and $ \pi_s^* $ (where $ 0 \leq s \leq T$) is an optimal dynamic hedging strategy in index $ S_t $. The pair $ (\alpha^*, \pi_s^*) $ should be obtained using a proper model. The same model should produce the highest/lowest price for which the trader should agree to buy or sell the $Z$-option.

In \cite{HalperinItkin2012} we developed a model that formalizes the above scenario by supplementing it with the specific dynamics for asset prices $S_t, Y_t $ and $ Z_t $, and providing criteria of optimality for pricing options $C_Z$. For the former, we use a standard correlated log-normal dynamics. For the latter, we employ the utility indifference framework with an exponential utility, pioneered by \cite{HN89,Davis97} and others, see e.g. \cite{HH2008} for a review. We showed that this results into a tractable setup with analytical (in quadratures) expressions for optimal hedges and option prices.
For more details and links to the related literature, see \cite{HalperinItkin2012}.

To extend this model, we notice that availability of just one asset for the static hedge in our
model is very restrictive. More generally, we may assume that $ N $ liquid options on
assets $Y_i, \ i  = 1, \ldots, N $ are available in the marketplace, where all $ Y_i $ have similar correlations $\rho_{z,y_i}$ with asset $ Z_t $. Therefore, all $ N $ options could be
used in this scenario to set up for static hedging of the $ Z$-option. All in total, we
have $ N + 1 $ assets for a static-dynamic hedge optimization problem.

This is the problem addressed by
the present work. Similar to  \cite{HalperinItkin2012}, we use the indifference pricing approach and an exponential utility function to derive a HJB equation for the value function.
In the present case, the HJB equation is $ (N+1)$-dimensional. We develop
an efficient numerical algorithm to solve the HJB equation.
Our approach is based on several changes of variables, a splitting scheme, and
a set of Fast Gauss Transforms (FGT), which turns out to be more efficient in terms of
complexity and lower local
space error than a finite-difference method.

%Therefore, it is more natural to consider a set of $N$ liquid assets $Y_i, \ i=1,N$ and
%options on them, as
%instruments available for hedging.

Before presenting our notation and convention, we note that while the mathematic framework developed below is general and can be applied to various
asset classes, for definiteness below we follow Ref. \cite{HalperinItkin2012} and specialize
on pricing and hedging on illiquid debt within a version of the Merton credit-equity model.
This setting might be of interest for modeling counterparty value adjustments (CVA) for over-the-counter (OTC) derivatives. However, a similar framework can be developed for other cases
where the utility-indifference approach is useful, e.g. executive stock options could
be priced along the same lines.

We therefore
assume a market where the following instruments can be traded:
\begin{itemize}
\item a risk-free zero-coupon bond $ B_0 $
\item a risky non-defaultable index $ S $
\item a set of liquid bonds $ B_{Y_i} $ issued by firms $ Y_i, \ i=1,N $, with market prices $ p_{Y_i}$
\end{itemize}
Our financial model for pricing and hedging an illiquid credit $ B_Z $ amounts to computing its price $ p_Z $ in terms of all $ p_{Y_i} $ and $ S$ at $t=0$, along with optimal hedges. Note that as long as issuers of $ Y_i $ and $ Z $ are imperfectly correlated, the liquid bonds $ B_{Y_i} $ provide only a partial hedge for $B_Z$.

As we are in the incomplete market setting, risk of $ Z$ cannot be perfectly hedged
by $ (B_{Y_i}, S ) $, hence both the price and hedge ratios will be different for different investors, depending on their risk preferences and a (non-unique) hedging strategy.

Therefore, the idea is to hedge an exposure to a counterparty with illiquid credit (a long position in bond $ B_Z $) by taking static short positions in a set of proxy liquid debts $B_{Y_i} $, plus possibly using a dynamic trading strategy $ \theta_t $ in the index $ S$.

Assume we statically hedge bond $ B_Z $ by selling $ \alpha_i $ zero-coupon bonds issued by firm $ Y_i $ for their market price $ p_{Y_i}$. The cash amount available for investing in bonds and index is $ x + \sum_i \alpha_i p_{Y_i} $, where $ x $ is the initial cash minus the price paid for $ B_Z $.

We further use an indifference pricing principle to derive an HJB equation which describes an evolution of the investor utility function in our setup.
As no closed form solutions are known for the utility indifference pricing with $N > 2$, we
suggest a very efficient numerical method of solving the HJB using a combination of a special change of variables and a particular splitting scheme.  Its total complexity is $O(M+N)M + O(M+1) \approx M^2, \ M \gg N$, where $M$ is the number of nodes for Fast Gauss Transform (FGT) used in calculations. This is significantly less
than e.g. the total complexity $O(N^M)$ of finite difference methods.
Both the theoretical setup and numerical algorithms presented below are the main results of
this paper, which to our knowledge are new.

The rest of the paper is organized as follows. Section~\ref{Framework} describes a general setup of the problem and indifference pricing framework. In section~~\ref{sHJB} we derive a corresponding HJB equation for our model. Section~\ref{sFV} introduces new factorized (adiabatic) variables, and shows that in new variables the HJB equation transforms to a $N$-dimensional heat equation with an extra non-linear term. This term is proportional to $\Phi^2_{y_0}/\Phi$, i.e. it contains only the first derivative of the dependent variable $\Phi$ wrt the first independent variable $y_0$ but no other derivatives. We describe an efficient  numerical algorithm to compute coefficients of such a transformation. The next section shows how the transformed HJB equation can be solved numerically using Strang's splitting. We show that the problem reduces to the solution of one $N$-dimensional and two one-dimensional heat equations.
For the latter task, we show how to use FGT to decrease the total complexity of the method.
Section~\ref{sC} discusses calibration of the method to the market data. The final section concludes.

%%%%%%%%%%%%%%%%%%%%%%%%%%%%%%%%%%%%%%%%%%%%%%%%%%%%%%%%%%%%%%%%
\section{Static hedging in indifference pricing framework} \label{Framework}
%%%%%%%%%%%%%%%%%%%%%%%%%%%%%%%%%%%%%%%%%%%%%%%%%%%%%%%%%%%%%%%%%
Borrowing from an approach of \cite{IlhanSircar2006} for a similar (but not identical) setting, we now show how the method of indifference utility pricing can be generalized to incorporate our scenario of a mixed dynamic-static hedge.

To this end, let $\Pi(Y_T, Z_T)$ be the final payoff of the portfolio consisting of our option positions, i.e.
\begin{equation} \label{Payoff}
\Pi^{\alpha}(Y_T, Z_T) = G_Z - \sum_i \alpha_i G_{Y_i}
\end{equation}

For convenience let us further denote asset $Z$ as $Y_0$. As long as all European options $C_{Y_i}, \ i \in [0,N]$ pay at the same maturity $T$, we can view this as the payoff of a combined ("static hedge portfolio") option $g(\alpha_0,...,\alpha_N)$, which involves payoffs $G_{Y_i}$ of all derivatives $C_{Y_i}$. Such option may be priced using the standard utility indifference principle. The latter states that the derivative price
$g(\alpha_0,...,\alpha_N)$ is such that the investor should be indifferent to the choice between two investment strategies. With the first strategy, the investor adds the derivatives to her portfolio of bonds and stocks
(or indices\footnote{The stock is equivalent to our index $S$ in the setting of the Merton's optimal investment problem.}) $S$, thus taking $g(\alpha_0,...,\alpha_N)$ from, and adding $\sum_i \alpha_i p_{Y_I} $ to her initial cash $x$. With the second strategy, the investor stays with the optimal portfolio containing bonds and the stocks/indices.

The value of each investment is measured in terms of the {\it value function} defined as the conditional expectation of utility $U(W_T)$ of the terminal wealth $W_T$ optimized over trading strategies. In this work, we use an exponential utility function
\begin{equation} \label{expUtility}
U(W) = - e^{ - \gamma W}
\end{equation}
\noindent where $\gamma$ is a risk-aversion parameter. In our case, the terminal wealth is given by
the following expression:
\begin{equation*}
W_T = X_T + \Pi^{\alpha_0,...,\alpha_N} (Y_{0,T},...,Y_{N,T})
\end{equation*}
\noindent with $X_T$ be the total wealth at time $T$ in bonds and index $ S $. In turn, the value function reads
\begin{align} \label{value_function}
V(&t,x,y_0,...y_N) = \\
&\sup_{ \pi_t \in \mathcal{M} } \mathbb{E} \left[ U \left(
 X_T + \Pi^{\alpha_0,...,\alpha_N} (Y_{0,T},...,Y_{N,T}) \right) \Big| X_t = x, Y_{0,t} = y_0, ...,
Y_{N,t} = y_N\right] \nonumber
\end{align}
\noindent where $ \mathcal{M} $ is a set of admissible trading strategies that require holding of initial cash $x$. The expectation in the \eqref{value_function} is taken under the ``real-world''
measure $\mathds{P}$.

For a portfolio made exclusively of stocks/indices and bonds, the value function for the exponential utility is known from the classical Merton's work:
\begin{equation} \label{Merton_vv}
V^0(x,t) = - e^{ - \gamma x e^{r \tau} - \frac{1}{2} \eta_s^2 \tau}
\end{equation}
\noindent where $\tau = T - t$, $r$ is the risk free interest rate assumed to be constant, and $\eta_s =
(\mu_s - r)/\sigma_s $ is the stock Sharpe ratio.

In our setting, in addition to bonds and stocks/indices, we want to long $C_{Y_0}$ option and short $\alpha_i$ units of every $C_{Y_I}$ option to statically hedge our $C_{Y_0}$ position, or, equivalently, buy the $g(\alpha_0,...,\alpha_N)$ option.

From the \eqref{value_function} the value function in our problem of optimal investment in bonds, index and the composite option $g(\alpha_0,...,\alpha_N)$ has the following form:
\begin{align} \label{valueFunction}
V(&t,x,y_0,...y_N) = \\
&\sup_{\pi_t \in \mathcal{M}} \mathbb{E}\left[ - e^{-\gamma
( X_T + \Pi^{\alpha_0,...,\alpha_N} (Y_{0,T},...,Y_{N,T}))}  \Big| X_t = x, Y_{0,t} = y_0, ...,
Y_{N,t} = y_N\right] \nonumber
\end{align}
\noindent where $X_T$ is a cash equivalent of the total wealth in bonds and the index at time $T$.
We represent it in a form similar to \eqref{Merton_vv}:
\begin{equation} \label{vv_HIM}
V(t,x,y_0,...y_N) = - e^{ - \gamma x e^{r \tau}- \frac{1}{2} \eta_s^2 \tau} \Phi(\tau, y_0,...y_N)
\end{equation}
\noindent where function $\Phi$ will be calculated in the next sections. The indifference pricing equation reads
\begin{equation*}
V(t,x,y_0,...y_N) = V^0\left(t,x + g(\alpha_0,...,\alpha_N) - \sum_{i=1}^N \alpha_i p_{Y,i}\right)
\end{equation*}
Plugging this in \eqref{Merton_vv} and \eqref{vv_HIM} and re-arranging terms, we obtain
\begin{equation*}
g(\alpha_0,...,\alpha_N) = - \frac{1}{\gamma} e^{ r \tau} \log \Phi(\tau, y_0,...y_N) + \sum_{i=1}^N \alpha_i p_{Y,i}
\end{equation*}
The highest price of the $Y_0$-derivative is given by choosing the optimal static hedge given by the numbers $\alpha_1,...,\alpha_N$ of the $Y_i$-derivatives, i.e.
\begin{align} \label{indif_price_2}
g(\alpha^*_0,...,\alpha^*_N) &= - \frac{1}{\gamma} e^{r \tau} \log \Phi^{(\alpha^*_0,...,\alpha^*_N)}(\tau, y_0,...y_N) + \sum_{i=1}^N \alpha_i p_{Y,i} \\
(\alpha^*_0,...,\alpha^*_N)  &= \arg \max_{\alpha^*_0,...,\alpha^*_N }  g(\alpha^*_0,...,\alpha^*_N) \nonumber
\end{align}
\noindent where we temporarily introduced a superscripts $\alpha_i$ in $\Phi^{(\alpha^*_0,...,\alpha^*_N)}$ to emphasize that the value function depends on all $\alpha_i$ through a terminal condition.

%==================================================
\section{The HJB equation} \label{sHJB}
%==================================================
To use the \eqref{indif_price_2} and thus be able to compute both the option price and optimal static hedge, we need to find the "reduced" value function $\Phi$. To accomplish this goal below we first derive the Hamilton-Jacobi-Bellman (HJB) equation for our model, and in the next section show how to efficiently solve it numerically.

Let $\theta$ be the investment strategy in the index. Optimal dynamic strategy can be obtained by using a general HJB principle
\begin{equation} \label{HJB}
V_t + \sup_{\pi} \mathcal{L}^\pi V = 0
\end{equation}
\noindent where $\mathcal{L}^\pi$ is the Markov generator, and $\pi = \pi_t(x)$ is the dynamic strategy at time $t$ which depends on the initial cash amount $x$.

Further assume that all state variables $S_t, Y_i, i \in [0,N]$ follow a geometric Brownian motion process with time-dependent drifts $\mu_i(t)$ and volatilities $\sigma_i, i \in (x,0,...,N)$
\begin{align*}
d S_t &= \mu_x(t) S_t dt + \sigma_x S_t d W_t^{(x)} \\
d Y_i &= \mu_i Y_i dt + \sigma_i Y_t d W_t^{(y)}, \quad i \in [0,N]
\end{align*}
Also following \cite{MusielaZariphopoulou2004} assume that a riskless bond $B_t = 1$ with maturity $T$ is available for
trading, yielding a constant interest rate $r$. Since our trading strategy implies a static position in all derivatives and dynamic positions in the index, real trading occurs in the time horizon $[t, T], 0 \le t \le T$, and only between the two traded assets, i.e., the riskless bond $B_t$ and the risky asset $S_t$. If our total wealth at time $ t $ is $ X_t = x $ and we invest amount $ \pi $ of this wealth into the index and the rest in a risk-free bond, the stochastic differential equation for $ X_t $ is obtained as follows:
\begin{equation*}
dX_t = r \left( X_t - \pi \right) dt + \frac{\pi}{S_t} dS_t = \left( r X_t + \pi \sigma_x \eta_s \right)
dt + \pi \sigma_x dW_t^{(x)} \; , \quad \eta_x = \frac{\mu_x -r}{\sigma_x}
\end{equation*}
Then $\mathcal{L}^\pi$ reads
\begin{align*}
\mathcal{L}^\pi &=  \left[ r x + \pi (\mu_x - r) \right] V_x
+ \frac{1}{2}\sigma^2_x \pi^2 V_{xx} + \sum_{i=0}^N \rho_{x,y_i} \sigma_x \sigma_{y_i} \pi y_i V_{x,y_i} \\
&+ \sum_{i=0}^N \mu_i  y_i V_{y_i} + \frac{1}{2}\sum_{i=0}^N \sum_{j=0}^N \rho_{ij} \sigma_i \sigma_j y_i y_j V_{y_{i},{y_j}}
\nonumber
\end{align*}
\noindent where $V(t,x,y_0,...y_N)$ is defined on the domain $\mathbb{R}(t,x,y_0,...,y_N): [0,T] \times [0, \infty) \times [0, \infty) \times ... \times [0, \infty)$.

Since $\mathcal{L}^\pi$ is a regular function of $\pi$, $\sup_{\pi}$ is achieved at
\begin{equation*}
\pi^*(x) =  - \frac{\eta_x V_x + \sum_{i=0}^N \rho_{x{y_i}} \sigma_i y_i V_{x{y_i}}}{V_{xx}}.
\end{equation*}
Plugging this into \eqref{HJB}, we obtain
\begin{align} \label{PDE}
V_t &+ r x V_x + \sum_{i=0}^N \mu_i  y_i V_{y_i} + \frac{1}{2}\sum_{i=0}^N \sum_{j=0}^N \rho_{ij} \sigma_i \sigma_j y_i y_j V_{y_{i},{y_j}} \\
&- \frac{1}{2} \frac{ (\eta_x V_x + \sum_{i=0}^N \rho_{xy_i} \sigma_{y_i} y_i V_{xy_i})^2}{V_{xx}} = 0
\nonumber
\end{align}
This is a nonlinear PDE with respect to the dependent variable $V(t,x,y_0,...,y_N)$ with standard boundary conditions (see \cite{MusielaZariphopoulou2004}), and the terminal condition determined by a choice of the writer's maximal expected utility (value function) of the terminal wealth $W_T$.

Note that so far the derivation is valid for a generic utility function. To make further progress, we specialize to the case of exponential utility in \eqref{expUtility}.
The latter choice gives rise to a natural dimension reduction of the HJB equation. Indeed, the ansatz
\begin{equation} \label{dimRed}
V(t,x,y_0,...,y_N) = - \exp\left(-\gamma x e^{r(T-\tau)}\right) G(\tau,z_0,...,z_N)
\end{equation}
\noindent with $z_i = \log (y_i/K_i), \ i \in [0,N]$ is both consistent with terminal condition \eqref{valueFunction}
and, upon substitution in (\ref{PDE}), leads to a PDE for function $G$ which does not contain variable $ x $:
\begin{equation} \label{GpdeWS}
    G_{\tau} =  - \frac{1}{2}\eta_x G  + \sum_{i=0}^N \hat{\mu}_i  G_{y_i} + \frac{1}{2}\sum_{i=0}^N \sum_{j=0}^N \rho_{ij} \sigma_i \sigma_j G_{y_{i},{y_j}}
    - \frac{1}{2 G} \left(\sum_{i=0}^N \rho_{x{y_i}} \sigma_i G_{y_{i}}\right)^2,
\end{equation}
\noindent where $\hat{\mu}_i = \mu_i - \frac{1}{2}\sigma^2_i - \eta_x \rho_{xy_{i}} \sigma_i$.

Equation \eqref{GpdeWS} is defined on the domain $\mathbb{R}(t,z_0,...,z_N): [0,T] \times [-\infty, \infty) \times...\times [-\infty, \infty)$. The initial condition for this equation is obtained from \eqref{valueFunction}.

In what follows, we choose a specific payoff of the form \eqref{Payoff} with $\Pi_i = \min(Y_i,K_i), \ i \in [0,N]$
where $K_i$ are strikes. Then the terminal condition for $ G(\tau,z_0,...,z_N)$ reads
\begin{equation} \label{initCondWS}
 G(0,z_0,...,z_N) = \exp \left[ - \gamma \left( K_0 e^{z_0^{-}} - \sum_{i=1}^N \alpha_i K_i e^{z_i^{-}} \right) \right]
\end{equation}
\noindent where $ z_i^{-} = \min(z_i,0)$.

%====================================================================
\section{The HJB equation and "factorized" variables} \label{sFV}
%====================================================================
The \eqref{GpdeWS} is a $(N+1)$-dimensional parabolic equation with a
 non-linear (quadratic) term.
 No closed form solution is available for this case. Note that when $ N = 1 $,
 the HJB equation can be solved using an asymptotic expansion proposed in \cite{HalperinItkin2012}. Another relevant reference is \cite{HendersonLiang2011} that studies a related problem of  counterparty risk of derivatives in incomplete markets with one traded and multiple non-traded assets\footnote{We note that our splitting method (see below) is different from that used by
the authors of \cite{HendersonLiang2011}. In addition, their method is of the first order in time, while
our method is of second order in time}.

Furthermore, straightforward applications of common numerical methods such as
e.g. finite differences would likely be inefficient in our setting. Indeed, assume that we approximate the non-linear term explicitly, as this does not affect stability of the FD scheme. \eqref{GpdeWS} then transforms to a $(N+1)$-dimensional linear parabolic equation with a source term, which would
be computationally costly to solve.

An alternative to this solution, yet straightforward numerical approach could be constructed as follows. We first use splitting
(see e.g. \cite{LanserVerwer}) that reduces the original $(N+1)$-dimensional problem to a set of
$N+1$ one-dimensional problems. Thus, if every one-dimensional grid contains $M$ nodes, and since every one-dimensional problem has a tridiagonal matrix, the total complexity of the method is
$O(M(N+1))$.
Next we use the Fast Gauss Transform \cite{IFGT} to solve the resulting one-dimensional
problems.

Below we show that this straightforward approach can be significantly improved by rewriting the \eqref{GpdeWS} in new "factorized" variables. The reason that we call these variable
"factorized" will be clear below.

First, make a change of the dependent variable $ G \rightarrow \Phi $ as follows:
\begin{equation} \label{changeOfFun1}
G(\tau,z_0,...,z_N) = e^{ - \frac{1}{2} \eta_x^2 \tau} \Phi(\tau,z_0,...,z_N),
\end{equation}
\noindent so the first term in the rhs of the \eqref{GpdeWS} drops off the equation for $\Phi$.

Our further idea is to build a map $y = (y_0...y_N) \rightarrow u = (u_0...u_N)$ such that in new variables $u$, both the Hessian matrix and the quadratic term in the \eqref{GpdeWS} become diagonal.

 To be more specific, let us first introduce some matrix notation. Let $A$ be the Hessian matrix, i.e. $A = \|\rho_{ij} \sigma_i \sigma_j\|, \ i,j \in [0,N]$. Let $a$ be a vector  $a = (\rho_{xy_i} \sigma_i), \ i \in [0,N]$. Let $R$ be a transformation matrix, i.e. $u = R^T y$. Then we want to find such $R$ that obeys
    \[ R^T A R = \lambda, \quad a R = B, \]
\noindent where $\lambda$ is some diagonal matrix, and $B = (1,0...0)$.

\begin{algorithm}
\caption{Algorithm of building matrix of transformations $R$.}
\begin{enumerate}
\item
Take a diagonal matrix $ D = \left\|
\begin{smallmatrix}
  d_2   & 0     & \cdots    & 0  & 0\\
  0     & d_3   & \cdots    & 0  & 0 \\
\vdots  & \vdots & \ddots   & \vdots & \vdots \\
  0     & 0     & \cdots & d_N & 0 \\
  0     & 0     & \cdots & 0 & -1  
\end{smallmatrix}
\right\|, $
\noindent where $d_2,...d_N$ are the unknowns to be determined.

\item Assign some initial values to $d_2,...d_N$ and solve an eigenvalues problem $R^{-1} D A R = \Lambda$, where $\Lambda$ is a diagonal matrix with eigenvalues at the diagonal. Then use the following proposition
\begin{proposition}  \label{Prop}
If $R$ is a matrix of eigenvectors of $D A$, e.g. $R^{-1} D A R = \Lambda$, then $R^T A R$ is a diagonal matrix.
\begin{proof}
See Appendix~\ref{A}
\end{proof}
\end{proposition}
\item Compute a vector $C = |a R - B|$. If all $C_2,...,C_N$ are less then the method tolerance $\varepsilon$ - we are done. Otherwise take the next guess on $d_2,...d_N$ and proceed until converge.
\end{enumerate}
\end{algorithm}

In other words to determine $d_2,..,d_N$ we have to solve a system of non-linear algebraic equations $a R = B$ with n = $2, \ldots,N$ wrt $d_2,...,d_N$, where matrix $R$ is defined implicitly via the solution of the eigenvalues problem $R^{-1} D A R = \Lambda$. This can be easily implemented, e.g. in Matlab just in few lines of code. The algorithm is pretty fast and converges to $\varepsilon = 10^{-15}$ within 25 msec for $N=4$ at Intel i7-2720 QM CPU 2.20 Ghz.

Since eigenvectors are defined up to scaling, we fix it by choosing value $-1$ in the right-bottom corner of matrix $D$ instead of adding an extra unknown $d_1$. Accordingly, we solve a system of $N-1$ equations $C_i = 0, \ i \in [2,N]$, rather then $N$ equations. This results in the fact, that the first element of vector $B$ could be whatever it becomes, rather than just 1.

Based on Proposition~\ref{Prop} we conclude that the above algorithm transforms the Hessian matrix to the diagonal form. At the same time the last step of the algorithm guarantees that in new variables the quadratic form in the nominator of the non-linear term in \eqref{GpdeWS} contains just one (namely, the first) term. That is exactly what we wanted to achieve by doing the proposed change of variables.

Some comments on the above algorithm should be made. First, logically the more our proxy assets correlate with the illiquid asset the better we can price the illiquid asset derivatives. This means that matrix $|\rho|$ has all elements, say in a range $0.5 \le |\rho_{ij}| \le 1$. Under these matrix $ D A$ becomes stiff with a high conditional number. Therefore, an accurate computation of eigenvalues and eigenfunctions of such a matrix requires high precision arithmetics. That means that at a 32-bit architecture the proposed algorithm could fail to converge to the true solution with the required accuracy (despite it converges to some solutions with a bigger error). Moving the algorithm to a 64-bit architecture significantly improved convergency but still could fail when $|\rho_{ij}|$ are close to 1. Therefore, in this case special algorithms of computing eigenvectors for stiff matrices have to be applied.

After the transformation matrix is found we finally use a change of independent variables
\[ u = R^T y + \tau M, \quad M = \left(\frac{1}{\tau} \int_0^\tau \hat{\mu}_0(k)dk, ..., \frac{1}{\tau} \int_0^\tau \hat{\mu}_N(k)dk \right), \]
\noindent to obtain
\begin{equation} \label{finEq}
\Phi_{\tau} = \frac{1}{2} \sum_{i=0}^N p_i \Phi_{y_{i},{y_i}}  - \frac{1}{2} b_0 \frac{\Phi^2_{y_0}}{\Phi}
\end{equation}
This can also be written as
\begin{align} \label{operEq}
\Phi_\tau &= \sum_{i=0}^N \mathcal{L}_i \Phi \\
\mathcal{L}_0 \Phi  = \frac{1}{2} p_0 \frac{\partial^2}{\partial y_0^2} \Phi - \frac{1}{2} b_0 \frac{\Phi^2_{y_0}}{\Phi},& \qquad
\mathcal{L}_i = \frac{1}{2} p_i \frac{\partial^2}{\partial y_i^2}, \ i=1,N \nonumber
\end{align}
Here $p_i, \ i \in [0,N]$ are the diagonal elements of the diagonal matrix $R^T A R$ (which is the Hessian matrix in new coordinates $u$), and $b_0$ is the first element of vector $B$.

It is seen that in new variables operators $\mathcal{L}_i, i \in [1,N]$ are linear. In addition, all operators
$\mathcal{L}_i, i \in [0,N]$ are independent. That is why we call these new variables $u$ as factorized.

\paragraph{Example.} Consider $N=3$ and the following parameters of the model:
\[
\rho_{y,y} = \left|
\begin{array}{cccc}
1.0 &  0.9 &  0.6 &  0.5 \\
0.9 &  1.0 &  0.75 & 0.7 \\
0.6 &  0.75 & 1.0 &  0.6 \\
0.5 &  0.7 &  0.6 &  1.0
\end{array}
\right|, \qquad
\begin{array}{ll}
\rho_{xy} = (0.23, 0.34, 0.45, 0.4), \\
\sigma_{y} = (0.3, 0.25, 0.35, 0.5)
\end{array}
\]
Use $d = (0.01,0.01,0.01,)$ as the initial guess. The above algorithm then produces the following solution:
\[
D = \left|
\begin{array}{cccc}
-0.06108 &  0 &  0 &  0 \\
0 &   0.2718 &  0 & 0 \\
0 &  0 & -0.1145 &  0 \\
0 &  0 &  0 &  -1
\end{array}
\right|, \qquad
R = \left|
\begin{array}{cccc}
-0.1180 &   0.2300 &   0.6490 &  -0.0990 \\
0.6466  & -0.9388  & -0.7556  &  0.2306 \\
-0.5047 &   0.1955 &   0.0117 &  -0.7905 \\
-0.5597 &   0.1657 &   0.0880 &   0.5587
\end{array}
\right|
\]

Accordingly, in the \eqref{finEq} $p = (0.0678, 0.0096, 0.0062, 0.0508)$ and $b_0 = -0.1446$. The total time of calculation is 0.6 sec on a 32 bit PC with 3.0 Ghz single core CPU.

%================================
\section{Numerical method} \label{NM}
%================================
To solve the \eqref{operEq} in general a $N$-th dimensional variant of Strang's splitting \cite{Strang} can be used which is $O(\Delta \tau^2)$. For linear operators this can be done by first formally solving the \eqref{operEq} in the form
\[ \Phi_\tau = \sum_i \mathcal{L}_i \Phi \ \rightarrow \ \Phi(\tau+\Delta \tau) = e^{ \Delta \tau \sum_i \mathcal{L}_i }\Phi(\tau) \]
\noindent and then applying a generalized BCH formula \cite{BCH}
\[
e^{ \Delta t \sum_i \mathcal{L}_i } = e^{ \frac{\Delta t}{2} \mathcal{L}_0} e^{ \frac{\Delta t}{2} \mathcal{L}_1}...e^{ \frac{\Delta t}{2} \mathcal{L}_{N-1}} e^{ \Delta t \mathcal{L}_N} e^{ \frac{\Delta t}{2} \mathcal{L}_{N-1}}...e^{ \frac{\Delta t}{2} \mathcal{L}_0}  + O(\Delta t^2) \qquad \
\]
For non-linear operators the situation is more delicate. However, as shown in \cite{ThalhammerKoch2010} the previous formal representation of the solution keeps to be valid in the non-linear case as well. Therefore, we can represent the previous equation as
\[ \Phi_\tau = \sum_i \mathcal{L}_i \Phi = \mathcal{L}_0 \Phi  + \mathcal{L}_{1N} \Phi,
\quad  \mathcal{L}_{1N} = \sum_{i=1}^N \mathcal{L}_i, \]
\noindent and use the Strang's splitting. Explicitly this means that at each time step we have to solve a system of three equations
\begin{align} \label{Strang}
\Phi^*_\theta &= \frac{1}{2} a_0 \frac{\partial^2}{\partial y_0^2} \Phi^* - \frac{1}{2} b_0 \frac{\Phi^{*,2}_{y_0}}{\Phi^*},
&\theta \in [0, \Delta \tau/2], \\
\Phi^{**}_\theta &= \mathcal{L}_{1N} \Phi^{**}, &\theta \in [0, \Delta \tau] \nonumber \\
\Phi^{***}_\theta &= \frac{1}{2} a_0 \frac{\partial^2}{\partial y_0^2} \Phi^{***} - \frac{1}{2} b_0 \frac{\Phi^{***,2}_{y_0}}{\Phi^{***}}, &\theta \in [0, \Delta \tau/2]
\nonumber
\end{align}
\noindent with the initial conditions for the first equation in \eqref{Strang}: $\Phi^*(0) = \Phi(\tau)$, for the second one: $\Phi^{**}(0) = \Phi^*(\tau+\Delta \tau/2)$, and for the last one: $\Phi(0) = \Phi^{**}(\tau + \Delta \tau)$. The final solution after this step is $\Phi(\tau + \Delta\tau) = \Phi^{***}(\tau + \Delta \tau/2)$.

Since our terminal condition is of a rather complicated form given in the \eqref{initCondWS}, all equations in
\eqref{Strang} can not be solved analytically, despite they do can be solved in quadratures. Indeed, the second equation is a $N$-dimensional heat equation which admits an efficient numerical solution by using Fast Gauss Transform (FGT) since the Green's function is this case is a $N$-dimensional Gaussian. The remaining equations by change of variables known as Cole-Hopf transformation \cite{ColeHopf}
\[
\bar{\theta} = a_0\theta, \ \bar{\Phi} = \Phi^{\frac{1}{1-(b_0/a_0)}}
\]
also reduces to the heat equation
\[
\bar{\Phi}_{\bar{\theta}} = \frac{1}{2}\bar{\Phi}_{y_0,y_0}
\]
Therefore, they also can be solved by using FGT.

Since we don't assume $N$ to be high, computation of the low-dimensional FGT doesn't face any difficulties if we use a powerful algorithm knows as Improved Fast Gauss Transform (IFGT) \cite{IFGT}. Consider first a one-dimensional heat equation. Its solution can be represented as a convolution of the initial condition with the Green's function (which in this case is the Gaussian kernel). Suppose that the discretized space variable $y$ is defined at $M$ state nodes (source nodes).
If we need to obtain the solution just at one fixed value of $y_0$, then we have one target point in space. However, according to the nature of the splitting algorithm we must solve similar problems at every splitting step (at given time we have 3 steps), and at every time step (the number of time steps $J$ is determined as $J=T/\Delta \tau$). Therefore, to re-apply IFGT we need to use our target points as the initial points at the next step. Therefore, the number of the target points is also $M$. Then the total complexity of IFGT is $O(2M)$.

For $d$-dimensional problem the number of source and target points is $M^d$. The complexity of IFGT is $O(2 M^d p(d)$ where $f(d,p)$ is a polynomial function of $d$ and the number of terms in $d$-variate Taylor expansion truncated after order $p-1$. To compare with finite-difference algorithms that usually are of the second order in space, consider an example with $p$=4 which provides a third order approximation. Thus, the total complexity of one step in time using Strang's splitting is $2 M^5 (f(5,p) + 2 f(1,p))$. As shown in \cite{IFGT}, e.g. $f(5,4)$ = 56, $f(1,4)$ = 4.
Therefore, the complexity of the five-dimensional IFGT with $M^5$ source points and $M^5$ target points is about $128 M^5$.

This could be compared with an analogous complexity of the finite difference method used to solve a d-dimensional heat equation at the space grid of $M^d$ nodes. Since all one-dimensional diffusion operators commute, this problem is reduced to five sequential one-dimensional problems. Every such a problem has the remaining $M^{d-1}$ states in other directions as dummy parameters, which means that this problem has to be solved $M^{d-1}$ times for every unique set of the dummy parameters. Also suppose we solve every problem with $k$ steps in time ($k = \theta/\Delta \theta)$). Then the total complexity of the method is $O(M)$ (the complexity of the FD one-dimensional solver for the heat equation, usually is about $6 M$) times $k$ (the number of steps in time), times $M^{d-1}$  the number of the dummy variables) times $d$ - the number of split tasks) which is $6 k d M^d $. For $d=5$ this gives $30 k M^5$.  Therefore, at $k > 4$ IMGT is faster
\footnote{Note that $k$=4 is too small for any FD scheme to eliminate some additional errors produced by discontinuity in the first derivative of the payoff function.}. At the same time the IMGT local error is essentially lower. That, as we mentioned, is because the standard schemes use the second order approximation in space
\footnote{This produces a tri-diagonal matrix, and the total complexity of the solver is about $6 M$. Better approximations, e.g. using Pade schemes, lead to banded matrices, therefore the total complexity, while still linear, grows significantly (see, e.g. \cite{ItkinCarr2012Kinky}).}, while the IFGT accuracy is defined by the number $p$, and is substantially higher.

Accordingly, doing $J$ steps in time results in the total complexity of the IFGT method to be $2 J M^N [f(N,p) + 2 f(1,p)] = O(J M^N)$. The proposed algorithm preserves the second order of approximation in time.

%%%%%%%%%%%%%%%%%%%%%%%%%%%%%%%%%%%%%%%%%%
\section{Calibration} \label{sC}
%%%%%%%%%%%%%%%%%%%%%%%%%%%%%%%%%%%%%%%%%%
To make this model practical one has to clearly understand how to calibrate the model to the market data. Two problems have to be discussed in this context.

First we need to calibrate the risk-aversion parameter $\gamma$. Though this parameters
may be specific to each investor, we may want to calibrate the risk aversion value
to a "representative" investor implied by the market. This problem was considered in \cite{BenthGrothLindberg2010} within a stochastic volatility model with a positive non-Gaussian Ornstein-Uhlenbeck process. Similar to our setup, the authors price options using the utility
indifference with an exponential utility. The model is calibrated to historical returns, and
the implied risk aversion is found by numerically inverting the indifference pricing equation given observed option prices. Certainly, in this case the risk aversion is a function of $T$ and $K$, e.g. $\gamma = \gamma(T,K)$.

An immediate problem with this approach is that when asset $Y_0$ is illiquid, it is hard to build the implied distribution of returns from the historical data, or to calibrate parameters of stochastic volatility for this asset. Therefore, in \cite{BenthGrothLindberg2010} liquid stocks (namely, MSFT and Volvo) were investigated. The initial intuition of the authors was that since the stochastic volatility model explains the observed market returns rather well, the implied risk aversion has to be almost flat with respect to $T$ and $K$ of the options. Contrary to this intuition, it was found
that implied risk aversion exhibits a smile behavior for short dated options, which was interpreted as
issuers' fear of a market crash (in the case of the issuance of a put option).
In particular, for Volvo, using call option bid/ask prices from December 30, 2005, it was found that
risk aversion $\gamma$ varies from 0.1 to 0.01. It decreases when maturity increases from 1 month to 1 year, and also increases when $K$ grows from 280 to 460. For puts the opposite is true, and the range of $\gamma$ is from 0.3 to 0. Similar behavior was observed for Microsoft, but in this case
$\gamma$ reaches 10 for puts at $K$ = 15 and $T$=1 month. These results give an idea of
a range of the implied risk aversion parameters.
However, it doesn't address the above question of how to apply this approach to an illiquid asset.

For some asset classes there sometimes exist other ways to imply the market value of $\gamma$. For instance, for FX this problem is considered in \cite{Stojanovic2011}. An essential property of the FX market is the existence of cross-currency rules. For simple models of the underlyings (such as
e.g. Geometric Brownian Motion, which is also our setting as well), this allows one
to express $\gamma$ in the explicit form via parameters of the domestic and foreign assets (see \cite{Stojanovic2011}, Eq.~7.3.11). An example which uses monthly data for USD and GBP between December 31, 1985 and August 31, 2005, and DJI and FTSE as market representatives, gives an estimation $\gamma$=4.17. This implies that the choice $\gamma$=1 with a logarithmic utility function which is frequently used in the literature might not be very realistic. Note that a closed form expression for $\gamma$ is obtained in \cite{Stojanovic2011} for stochastic interest rates.

Another challenge closely related to the first problem consists of the fact that for the illiquid asset $Y_0$, it is hard to find its correlation with the potential candidates to be
the proxy assets, $Y_1,...,Y_N$, essentially almost by definition, as an illiquid asset
typically does not move enough to measure its correlation with other assets.
One way to proceed in such case is to use other, liquid assets from the same economic sector
as $ Y_0 $, as "correlation proxies", as a way to roughly calibrate correlation parameters
of $ Y_0 $ and our liquid proxies, which are the inputs in our framework.
Note that in order to serve as a credible "correlation proxy" for $ Y_0 $, another (liquid) proxy
$ Y_0' $ is expected to be similar to $ Y_0 $, e.g. they should have similar credit ratings, credit default swap (CDS) spreads, expected
default frequency (EDF) etc.

%Therefore, we face an unpleasant situation when having no a reliable value of the correlation a choice of a suitable proxy asset is pretty ambiguous. Other arguments then should be attracted such as i) the proxy asset should belong to the same economical sector as $Y_0$; ii) two corporations that issued assets $Y_0$ and $Y_i$ have the same creditworthiness and credit rating; etc. And to better understand how to proceed with this clarification is necessary of what we actually suspect by the illiquid asset.

The following differences of an illiquid asset from its liquid counterpart is discussed in \cite{APW2011}. First, an illiquid asset $Y_0$ can only be rebalanced at infrequent, stochastic intervals. When a trading opportunity arrives, the investor is able to rebalance her holdings of the illiquid asset. Furthermore, an illiquid asset is an asset that is not traded in a centralized exchange. In this case, investors who are willing to trade in this asset need to search for a counterparty. Such search process might be time-consuming, since in many cases the number of market participants with the required expertise, capital, and interest in these illiquid assets could be small. Examples of such illiquid assets are hedge funds, venture capital, private equity, structured credit, and real estate. Some of these assets are traded in OTC markets, but in others investors need to search directly for a counterparty in order to rebalance a position.

The second way in which the illiquid asset differs from the liquid assets is that it cannot be pledged as collateral. Investors can issue non-state contingent debt by taking a short position in the riskless bond, but they cannot issue risky debt using the illiquid asset as collateral. If investors were allowed to do so, they could convert the illiquid asset into liquid wealth, and thus would implicitly circumvent the illiquidity friction.

This analysis means that the correlation between the illiquid asset $Y_0$ and other proxy assets, at least in principal, can be computed from historical data, referring either $ Y_0 $ (or its "correlation proxy" asset $ Y_0' $). However, this is a delicate issue since the historical times series for $Y_0$ are recorded with time periods demonstrating kind of stochastic behavior. From this prospective an extended Kalman filter is a proper tool to work with the sparse, irregular time series. For more detail, see, e.g. \cite{Gleb1974}. Another prominent approach is a spectral estimation of a non-stationary time series sampled with missing data. The time series could be modeled as a locally stationary wavelet process, and its realization is assumed to feature missing observations \cite{KnightNunesNason2012}.

%%%%%%%%%%%%%%%%%%%%%%%%%%%%%%%%%%%%%%%%%%%%%%%
\section{Conclusions}
%%%%%%%%%%%%%%%%%%%%%%%%%%%%%%%%%%%%%%%%%%%%%%%
In this paper we proposed a framework for pricing derivatives written on illiquid asset using a mixed dynamic-static hedging in a proxy index and $N$ proxy options. While in this paper we apply our framework to an incomplete market version of the credit-equity Merton's model, the same approach can be used for other asset classes (equity, commodity, FX, etc.), e.g. for pricing and hedging options with illiquid strikes or illiquid exotic options, executive stock options etc.

An efficient numerical algorithm is proposed which combines several changes of independent variables at the first step and Strang's splitting at the second step.

A linear change of variables to new factorized (adiabatic) variables transforms the HJB equation for our model into a $(N+1)$-dimensional heat equation with an extra non-linear term. This term is proportional to $\Phi^2_{y_0}/\Phi$, i.e. it contains only the first derivative of the dependent variable $\Phi$ wrt the first independent variable $y_0$. This in contrast to the original HJB equation that has mixed derivatives, drifts and the non-linear term of the form $(\sum_{i=0}^N \Phi_{y_0})^2/\Phi$. We
propose an efficient  numerical algorithm to compute coefficients of this linear transform. Some peculiarities of the algorithm are discussed. In particular, in the case of strong correlations between the illiquid asset $Y_0$ and other proxy assets $Y_1,...,Y_N$, the diagonal matrix $ D $ which we have to
compute
%(defined by the relation $D A = R \Lambda R^{-1}$, where $A_{ij} = \rho_{ij}\sigma_i \sigma_j$, $R$ is the transformation matrix and $\Lambda$ is the eigenvalues matrix of $D A$)
could be stiff. In this case, computation of eigenvectors of a non-symmetric matrix $D A$ could require special methods (preconditioners) to preserve accuracy of computations.

At the next step this new HJB equation in new variables is solved numerically using Strang's splitting. We show that this problem reduces to the solution of one $N$-dimensional and two one-dimensional heat equations. Furthermore, we  propose to use the Improved Fast Gauss Transform to
decrease the total complexity of the method. We demonstrate that this complexity is $2 J M^N [f(N,p) + 2 f(1,p)]$, where $J$ is the number of steps in time, $M$ is the number of grid points in $S$, and function $f(m,n)$ is defined in \cite{IFGT}.
This algorithm is of the second order of approximation in time and of the $p-1$ order of approximation in space. We also compare this with the finite-difference algorithm and show that our proposed algorithm produces less error and is more efficient in performance.

In this paper for all assets we used a GBM model with time-dependent drifts and constant volatilities $\sigma_i$. But this approach can also be generalized when volatilities $\sigma_i = \sigma_i(t)$ are functions of time. This case will be discussed elsewhere.

%==================================================
\section*{Acknowledgments}
We thank Peter Carr, Alex Lipton, and attendees of the "Global Derivatives USA 2012" conference for useful comments. I.H. would like to thank Andrew Abrahams and Julia Chislenko for support and interest in this work. We assume full responsibility for any remaining errors.

%==================================================
\clearpage
%\def\myBib{C:/AndreyItkin/MySettings2011/aitkin_fin}
%\def\myBib{C:/NxData/AItkin/Papers/MySettings2011/aitkin_fin}
%\bibliographystyle{plain}
%\bibliography{\myBib}
\newcommand{\noopsort}[1]{} \newcommand{\printfirst}[2]{#1}
  \newcommand{\singleletter}[1]{#1} \newcommand{\switchargs}[2]{#2#1}

%%%%%%%%%%%%%%%%%%%%%%%%%%%%%%%%%%%%%%%%%%%%%%%%%%

\newpage
\begin{appendices}
\markboth{Appendices}{Appendices}

%%%%%%%%%%%%%%%%%%%%%%%%%%%%%%%%%%%%%%%%%%%%%%%%%%%%%%%%%%%%%%%%%%%%%%%%%%%%%
\section{Proof of Proposition \ref{Prop}} \label{A}
%%%%%%%%%%%%%%%%%%%%%%%%%%%%%%%%%%%%%%%%%%%%%%%%%%%%%%%%%%%%%%%%%%%%%%%%%%%%%
Proposition~\ref{Prop} claims that given a diagonal matrix $D$ and a symmetric real matrix $A$, and
matrix $R$ of eigenvectors of $D A$, e.g. $R^{-1} D A R = \Lambda$, where $\Lambda$ is a diagonal matrix with eigenvalues at the diagonal, it follows that $R^T A R$ is a diagonal matrix.
\begin{proof}
By definition $D A = R \Lambda R^{-1}$. Multiply both sides of this expression by $D^{-1/2}$ from the left and by $D^{1/2}$ from the right to obtain
\[ D^{1/2} A D^{1/2} = D^{-1/2} R \Lambda R^{-1} D^{1/2} \]
Introducing a diagonal matrix $\Xi$ such that $\Xi \Xi^{-1} = I$ - a unit matrix, this can also be rewritten as
\begin{equation} \label{p1}
D^{1/2} A D^{1/2} = D^{-1/2} R \Xi \Xi^{-1} \Lambda  R^{-1} D^{1/2} = D^{-1/2} R \Xi \Lambda \Xi^{-1} R^{-1} D^{1/2}.
\end{equation}
Matrix $D^{1/2} A D^{1/2}$ is a symmetric complex matrix, therefore it can be decomposed using its eigenvectors $\bar{R}$ and eigenvalues $\Lambda$ which coincide with that of the matrix $D A$.
\begin{equation} \label{p2}
D^{1/2} A D^{1/2} = \bar{R} \Lambda \bar{R}^{-1}
\end{equation}
Comparing the \eqref{p1} and \eqref{p2} we see that eigenvectors $\bar{R}$ and $R$ are connected by the map
\begin{equation} \label{map}
R = D^{1/2} \bar{R} \Xi^{-1}
\end{equation}
Using this map and taking into account that $D$ and $\Xi$ are diagonal matrices we can transform the matrix $R^T A R$ as follows
\begin{align*}
R^T A R &= [D^{1/2} \bar{R} \Xi^{-1}]^T A D^{1/2} \bar{R} \Xi^{-1} = \Xi^{-1} \bar{R}^T (D^{1/2})^T A D^{1/2} \bar{R} \Xi^{-1} \\
&= \Xi^{-1} \bar{R}^{-1} D^{1/2} A D^{1/2} \bar{R} \Xi^{-1} = \Xi^{-1} \Lambda \Xi^{-1}.
\nonumber
\end{align*}
Here we used the fact that the matrix $D^{1/2}$ is diagonal; $D^{1/2} A D^{1/2}$ is a complex symmetric matrix, therefore its eigenvectors $\bar{R}$ are orthogonal and $\bar{R}^T = \bar{R}^{-1}$.

The last step of the proof is to recognize that since matrices $\Lambda$ and $\Xi$ are diagonal, the product $\Xi^{-1} \Lambda \Xi^{-1}$ is a diagonal matrix as well.

Note, that matrix $\Xi$ is not an arbitrary matrix. It is determined by the \eqref{map} and is
\[ \Xi = R^{-1} D^{1/2} \bar{R} \]
\noindent Accordingly,
\[ R^T A R = \bar{R}^{-1} D^{-1/2} R \Lambda \bar{R}^{-1} D^{-1/2} R \]

\end{proof}
\end{appendices}
\end{document}